\documentclass[aps,12pt,axodraw]{article}
\usepackage{epsfig,sint,macros}

\begin{document}

\begin{titlepage}
\begin{flushright}
MITP/13-050
\end{flushright}

\renewcommand{\thefootnote}{\fnsymbol{footnote}}

\vskip 0.5 cm
\begin{center}
  {\Large\bf 
New chiral lattice actions of the Bori\c{c}i-Creutz type
\\[0.5ex]}
\end{center}
\vskip 0.5 cm
\begin{center}
{\large Stefano Capitani\footnote{capitan@kph.uni-mainz.de}}
\vskip 0.5cm
Institut f\"ur Kernphysik and HIM (Helmholtz-Institut Mainz), \\
University of Mainz, Johann-Joachim-Becher-Weg 45, D-55099 Mainz, Germany
\vskip 1.0cm
{\bf Abstract}
\vskip 0.35ex
\end{center}

\renewcommand{\thefootnote}{\arabic{footnote}}

\noindent
We generalize the Bori\c{c}i-Creutz action in such a way that the position
of the second zero and the direction which breaks the hypercubic symmetry can
be arbitrarily chosen, and the action has still the correct continuum limit.
Minimal doubling is guaranteed if the distance between the two zeros does 
not become too large. Special values of this distance could turn out to be
particularly convenient for efficient numerical simulations of minimally
doubled fermions.
\vfill

\begin{center}
November 2013
\end{center}

\eject

\vfill
\eject

\end{titlepage}

\setcounter{footnote}{0}

\section{Introduction}

Graphene-inspired Bori\c{c}i-Creutz fermions
\cite{Creutz:2007af,Borici:2007kz,Creutz:2008sr,Borici:2008ym} 
have sparked off a few years ago a revival of minimally doubled actions
(a recent overview is in \cite{Misumi:2012eh}). With these simple lattice
formulations Monte Carlo simulations of two degenerate flavors preserving
chiral symmetry for any finite lattice spacing $a$ can be carried out.
Two flavors is the minimum value allowed by the Nielsen-Ninomiya theorem
if one wants to keep an exact continuous chiral symmetry (of the standard type,
i.e. not Ginsparg-Wilson) and also maintain other convenient field-theoretical
properties like locality and unitarity.

These chiral fermionic formulations can be still kept ultralocal (like Wilson
fermions), and since they contain only nearest-neighbor interactions, they are
cheap and easy to simulate. Their formalism is also quite simple, and one can
construct conserved axial currents which have a simple and compact expression.

Minimally doubled fermions have the same kind of $U(1)$ chiral symmetry as 
staggered fermions, and are slightly more expensive compared with them.
However, having 2 flavors instead of 4 they do not require uncontrolled
extrapolations to 2 physical light flavors, and so they are ideal for $N_f=2$
simulations. One also avoids the complicated intertwining of spin and flavor
of staggered fermions.\,\footnote{Very recently a minimally doubled staggered
action has been presented by Creutz in \cite{Creutz:2013ofa}. This action
contains hermitian as well as anti-hermitian parts, and breaks hypercubic
symmetry like all other minimally doubled actions found so far (in this case
it is the temporal direction which is treated in a different way).}

As they are much cheaper than Ginsparg-Wilson fermions, minimally doubled
fermions can be very convenient for vector-like theories like QCD. Moreover,
they might be very practical for simulations of lattice QCD at finite
temperature, where staggered fermions are extensively used.

Bori\c{c}i-Creutz fermions have a special place among minimally doubled
fermions, not only because they have sparked off the revival of this class of
ultralocal chiral formulations, but also for their particular construction
which has arisen from investigations of the properties of electrons in graphene.
Bori\c{c}i-Creutz fermions are an instructive example of models based on
spinless fermions hopping on a lattice, in which the low-energy excitations
come out at the end to carry half-integer spin.\,\footnote{Another example is
given by the above mentioned minimally doubled staggered fermions of Creutz
\cite{Creutz:2013ofa}, constructed out of a lattice of spinless fermions
subjected to a constant magnetic field.} On the lattice strictly speaking the
spin-statistics theorem is not valid (as relativistic invariance is broken),
and the emergence of spin from spinless particles has been put forward by
Creutz and recently discussed in \cite{Creutz:2013ofa}. How the spin arises is
dictated by the topological behavior of the action in momentum space. It is
quite interesting to see how the topological protection from additive mass
renormalization works, and also how the topological properties of the action
constrain the fermionic flavors to appear only in an even number, giving so an
intriguing picture of the workings of the Nielsen-Ninomiya theorem.

Together with Bori\c{c}i-Creutz fermions, another simple realization of
minimally doubled fermions known as Karsten-Wilczek fermions
\cite{Karsten:1981gd,Wilczek:1987kw} has also been studied in some detail in
the last few years. The minimal doubling of the latter formulation comes out
through a more ``standard'' mechanism, more similar to that of Wilson fermions.
Many properties of these two actions have been studied and more deeply
understood in
\cite{Capitani:2009yn,Capitani:2009ty,Capitani:2010nn,Capitani:2010ht},
where the Bori\c{c}i-Creutz and Karsten-Wilczek fermions have been shown
to form a fully consistent renormalized quantum field theory. 

The standard Karsten-Wilczek action, which has the two zeros at a fixed
location, has been recently generalized in such a way that an arbitrary
distance between its two zeros can be chosen
\cite{Creutz:2010qm,Creutz:2011hy,Capitani:2013zta}.
Achieving something similar for the standard Bori\c{c}i-Creutz action as well,
where the position of the second zero is also constrained to a specific
location, has turned out to be less straightforward. In this article we present
the first successful formulation of a generalized Bori\c{c}i-Creutz action with
variable distance between the two zeros. Moreover, the direction of hypercubic
breaking can also be chosen at will.

In general it can be convenient to have at hand minimally doubled actions where
the distance between the two poles of the quark propagator can be arbitrarily 
varied. Special values of this distance could also provide actions which are
more advantageous for numerical simulations, in that for instance they minimize
some artefacts specific to these formulations. Here we have a $U(1)$ chiral
symmetry instead of the continuum $SU(2)$, and only the neutral pion remains
as a Goldstone boson, while the charged pions will have a higher mass. The
effective amount of important physical quantities such as the mass difference
between the $\pi^\pm$ and the $\pi^0$, or of mass splittings within otherwise
degenerate multiplets, could turn out to be rather small for a few of these
actions and not so small for all other ones. Then, having the possibility of
moving the distance between the two poles could be useful in order to minimize
in the continuum limit the effects (among others) coming from having only a
$U(1)$ chiral symmetry.

Thus, it is always useful to possess as many different minimally doubled
actions as possible, and keep on trying to construct new ones. Some particular
actions could turn out to have better theoretical or practical properties, and
be particularly advantageous for lattice simulations of chiral fermions. 

This article is organized as follows. In Sect. \ref{sec:generalized} we
recall the form of the standard Bori\c{c}i-Creutz action and present its
generalization, which allows an arbitrary position of the second zero, while in 
Sect. \ref{sec:construction} we explain how one can arrive at this generalized
action. Then in Sect. \ref{sec:dirac} we give the new Dirac matrices that have
to be introduced, while in Sect. \ref{sec:minimal} we discuss which choices
of the position of the second zero allow minimal doubling to be preserved,
before giving an outlook in Sect. \ref{sec:outlook}. Finally, in Appendix
\ref{sec:derivations} we provide a few derivations of important properties of
the zeros, which make use of the trace equations.

\section{The generalized action}
\label{sec:generalized}

The work of Creutz \cite{Creutz:2007af,Creutz:2008sr} and Bori\c{c}i 
\cite{Borici:2007kz,Borici:2008ym} has led to the formulation of a fermionic
action whose Dirac operator in momentum space is given, in the free case, by
\begin{eqnarray}
D^{BC}(p) &=& 
\frac{i}{a} \,\sum_\mu \Big(\gamma_\mu \sin ap_\mu
- \gamma'_\mu (1-\cos ap_\mu)\Big) +m_0 
\nonumber \\
&=& \frac{i}{a} \, \sum_\mu \Big(\gamma_\mu \sin ap_\mu 
+ \gamma'_\mu \cos ap_\mu\Big) - \frac{2i\Gamma}{a} +m_0 ,
\label{eq:bc}
\end{eqnarray}
where 
\begin{equation}
\gamma'_\mu = \Gamma \gamma_\mu \Gamma
\end{equation}
and
\begin{equation}
\Gamma = \frac{1}{2} \, \sum_\mu \gamma_\mu = \frac{1}{2} \, \sum_\mu \gamma'_\mu,
\label{eq:bcbiggamma}
\end{equation}
with $\Gamma^2=1$. This action vanishes at $ap_\mu=(0,0,0,0)$ and 
$ap_\mu=(\pi/2,\pi/2,\pi/2,\pi/2)$ and describes two fermion species of
opposite chirality. The two zeros determine the special direction which breaks
the hypercubic symmetry, which is in this case a major hypercubic diagonal.
This special direction is also encoded in the expression of the matrix
$\Gamma$.

We generalize here the Bori\c{c}i-Creutz action in such a way that the second
zero can be moved to an arbitrary position $ap_\mu=\alpha_\mu$, with
$-\pi < \alpha_\mu < \pi$ (but $\alpha_\mu \neq 0$). The components of
$\alpha_\mu$ do not need to be equal, and actually they can even be all
different from one another. Thus, the direction of hypercubic breaking can be
arbitrarily chosen. However, as we will understand later, this direction can
never exactly correspond to one of the $p_\mu$ axes.

The Dirac operator of the generalized Bori\c{c}i-Creutz action that we have
found is given in position space by
\begin{equation}
D = \frac{1}{2} \, \Bigg\{ 
\sum_{\mu=1}^4 \gamma_\mu (\nabla_\mu + \nabla^\ast_\mu) \, 
+ia \sum_{\mu=1}^4 \Big(
\gamma_\mu \cot \alpha_\mu + \gamma'_\mu \csc \alpha_\mu
\Big) \,\nabla^\ast_\mu \nabla_\mu \Bigg\} + m_0, 
\end{equation}
where 
\begin{equation}
\nabla_\mu\,\psi (x) = \frac{U_\mu(x) \psi(x+a\widehat{\mu}) - \psi (x)}{a}
\end{equation}
is a lattice discretization of the covariant derivative, and $\gamma'$ is
another set of Dirac matrices, which will be specified later. After expanding 
the covariant derivatives this fermionic action reads
\begin{eqnarray}
& & S^f = a^4 \sum_{x} \Bigg\{ \frac{1}{2a} \sum_{\mu=1}^4 \bigg[
    \overline{\psi} (x) \, \bigg(\gamma_\mu + i\,\Big(\gamma_\mu \cot \alpha_\mu 
+ \gamma'_\mu \csc \alpha_\mu \Big)\bigg) \, 
   U_\mu (x) \, \psi (x + a\widehat{\mu}) \nonumber \\
&& \phantom{S^f = a^4 \sum_{x} \Bigg\{ \frac{1}{2a} \sum_{\mu}} \quad
-\overline{\psi} (x + a\widehat{\mu}) \, \bigg(\gamma_\mu - i\,\Big(\gamma_\mu
  \cot \alpha_\mu + \gamma'_\mu \csc \alpha_\mu \Big)\bigg) \,
   U_\mu^\dagger (x) \, \psi (x) \bigg] \nonumber \\
&& \phantom{S^f = a^4 \sum_{x} \Bigg\{ \frac{1}{2a}} 
+ \overline{\psi}(x) \, \bigg( m_0-\frac{i}{a} \, \sum_\mu \, \Big(
\gamma_\mu \cot \alpha_\mu + \gamma'_\mu \csc \alpha_\mu \Big) \bigg) \, \psi (x) 
    \Bigg\} ,
\label{eq:generalized}
\end{eqnarray}
and it has, like the Wilson action, only nearest-neighbor interactions.
This action has the correct continuum limit, and since (as we will see later
in Sect. \ref{sec:dirac}) one also has $\{\gamma'_\mu,\gamma_5\}=0$, 
it preserves a $U(1)$ chiral symmetry (for $m_0=0$), which protects from
additive mass renormalization, and also satisfies $\gamma_5$-hermiticity.

\section{Construction of the generalized action}
\label{sec:construction}

We now explain how the generalized action (\ref{eq:generalized}) can be
derived. We carry out our reasoning in momentum space, and it is sufficient to
consider the free case and then at the end insert the gauge interactions.
From now on we will set for simplicity $a=1$.

The standard Bori\c{c}i-Creutz action (\ref{eq:bc}) can be viewed as the
outcome of an ingenious construction, which was devised by Creutz in
\cite{Creutz:2008sr}. It can be represented as a linear combination of two
physically equivalent naive fermion actions (one of them having been given
a momentum shift), which are the first and second term in the last line of 
Eq.~(\ref{eq:bc}). The first term vanishes when any component of the momentum
is equal to $0$ or $\pi$. The zeros of the second term are instead positioned
at the momenta $(\pm \pi/2,\pm \pi/2,\pm \pi/2,\pm \pi/2)$, which are the ones
that are maximally distant from the zeros of the first term. So, the 16 zeros
corresponding to the doublers of the second naive action are located
farthest away from the 16 zeros of the doublers of the first naive action.
The particular combination in (\ref{eq:bc}), with the subtraction of $\Gamma$,
results in an action with only two zeros.

We first try to use again two naive fermions in order to arrive at a generalized
action in which the second zero can be put at an arbitrary location. To do this,
it is useful to see the Bori\c{c}i-Creutz action written in the following way:
\begin{equation}
D^{BC}(p) = i \,\sum_\mu \Big(\gamma_\mu \sin p_\mu 
+ \gamma'_\mu \sin (p_\mu+\pi/2)\Big) - 2i\Gamma +m_0 .
\end{equation}
This suggests to make a translation in momentum space of the second naive
fermion action, such that its second zero on each of its axes (the one with
the negative slope) is moved to a generic position $p_\mu=\alpha_\mu$:
\begin{equation}
D^{BC'}(p) = i \, \sum_\mu \Big(\gamma_\mu \sin p_\mu
 + \gamma'_\mu \sin (p_\mu+\pi-\alpha_\mu)\Big)
 -i\,\sum_\mu \gamma'_\mu \sin \alpha_\mu +m_0 .
\label{eq:action2} 
\end{equation}
Then the second zero of the minimally doubled action (\ref{eq:action2}) is at
$p_\mu=\alpha_\mu$ (the first one still sitting at the origin). Note that also
the $\Gamma$ term had to be modified in order to achieve the desired minimal
doubling, and is now $\Gamma=(1/2) \sum_\mu \gamma_\mu \sin \alpha_\mu = (1/2) 
\sum_\mu \gamma'_\mu \sin \alpha_\mu$, and that this action can also be written
as
\begin{eqnarray} 
D^{BC'}(p) &=& i \, \sum_\mu \Big(\gamma_\mu \sin p_\mu + \gamma'_\mu 
\big(\sin (\alpha_\mu-p_\mu) - \sin \alpha_\mu \big) \Big) +m_0 \nonumber \\
  &=& i \, \sum_\mu \Big( \gamma_\mu \big(\sin p_\mu - \sin \alpha_\mu \big)
 + \gamma'_\mu \sin (\alpha_\mu-p_\mu)  \Big) +m_0 \nonumber .
\end{eqnarray}

The mechanism out of which one at the end obtains an action like
(\ref{eq:action2}) which has only two zeros is analogous to the one of the
standard Bori\c{c}i-Creutz action. When the coefficients of both $\gamma_\mu$
and $\gamma'_\mu$ vanish at the same value of the momentum $p_\mu$, one has a
zero of the action (although this is not the only possibility, because
$\gamma_\mu$ and $\gamma'_\mu$ are not independent). This is true for any given
spacetime direction and the corresponding momentum component $\mu$. Leaving
aside for a moment the last term of the action (\ref{eq:action2}) (the one
containing $\sum_\mu \gamma'_\mu \sin \alpha_\mu$), we can see that the
coefficient of $\gamma_\mu$ in the first term and that of $\gamma'_\mu$ in the
second term can never vanish together. Moreover, the value of the coefficient of
$\gamma_\mu'$ at the momentum where the coefficient of $\gamma_\mu$ vanishes
with positive slope, is the same of the coefficient of $\gamma_\mu$ at the
momentum where the coefficient of $\gamma'_\mu$ vanishes with negative slope,
and is equal to $i\sin \alpha_\mu$. That is why one at the end needs to
subtract  the term with $\sum_\mu \gamma'_\mu \sin \alpha_\mu$ from the sum of
the two naive actions. The outcome of this construction is that only two zeros
are left, one at the origin and the other at $\alpha_\mu$.

There is however a major problem with the action (\ref{eq:action2}), and it is 
that it does not have the correct continuum limit. Indeed, its leading terms
for small $p$ are
\begin{equation}
D^{BC'}(p) \, \simeq \, i\slash{p} - i\sum_\mu \gamma'_\mu\,p_\mu \cos \alpha_\mu,
\label{eq:wrong-cl}
\end{equation}
and one consequence of this is that the basic vertex for the emission of a
gluon by a quark current is not simply proportional to $\gamma_\mu$, but
contains also $\gamma'_\mu$ terms, even in the continuum limit. Then this
action is not suitable to be used in Monte Carlo simulations.

The wrong continuum limit originates from the fact that at the point
$p_\mu=(0,0,0,0)$, where the coefficient of $i\gamma_\mu$ vanishes, the first
derivative of the function expressing the coefficient of $i\gamma'_\mu$ does not
vanish. One must then find a way to overcome this limitation.

In order to obtain that this derivative becomes zero, we have to modify in a
suitable way the shape of the naive actions in momentum space. This can be
accomplished by making the substitution
\begin{equation}
\sin p_\mu \, \longrightarrow \, \sin p_\mu - \cot \alpha_\mu \,(1 -\cos p_\mu) .
\label{eq:modfunction}
\end{equation}
This new function is now taken as the coefficient of $i\gamma_\mu$. It is
easy to see that on each $\mu$ axis it has zeros at $p_\mu=0$ and
$p_\mu=2\alpha_\mu$, with slopes $1$ and $-1$ respectively (as it should be).
Its maximum is attained at $p_\mu=\alpha_\mu$, with value
$(1-\cos \alpha_\mu)/\sin \alpha_\mu$, and its minimum at
$p_\mu=\alpha_\mu-\pi$, with value $-(1+\cos \alpha_\mu)/\sin \alpha_\mu$.
The coefficient of $i\gamma'_\mu$ is given by a function in momentum space 
of the same shape as (\ref{eq:modfunction}), but on which a negative shift
of length $\alpha_\mu$ has been applied. Its zeros are then located at
$-\alpha_\mu$ and $\alpha_\mu$, and the position of the maximum is also
shifted accordingly.

The mechanism of minimal doubling is now similar to the one sketched above.
The main difference is that at $p_\mu=0$, where the coefficient of $\gamma_\mu$
is zero, that of $\gamma'_\mu$ has now a maximum, and thus its first derivative
is zero. At $p_\mu=\alpha_\mu$ the roles of $\gamma_\mu$ and $\gamma'_\mu$ are
simply reversed. The (free) minimally doubled Dirac operator coming out of
this choice of modified naive actions is then
\begin{eqnarray}
D(p) &=& i \, \sum_\mu \bigg[ \gamma_\mu \,\Big( \sin p_\mu
 - \cot \alpha_\mu \,(1 -\cos p_\mu) \Big) \nonumber \\
&& \qquad \, + \, \gamma'_\mu \,\Big( \sin (p_\mu+\alpha_\mu) 
 - \cot \alpha_\mu \,(1 -\cos (p_\mu+\alpha_\mu) ) \Big) \bigg]
 - i n \Gamma +m_0 
\nonumber \\
&=&  i \, \sum_\mu \frac{1}{\sin \alpha_\mu} \bigg[
 \gamma_\mu \, \Big( \cos (p_\mu-\alpha_\mu) - \cos \alpha_\mu \Big)
 \nonumber \\
&& \qquad \qquad \quad + \, \gamma'_\mu \,\Big( \cos p_\mu - \cos \alpha_\mu
 \Big) \bigg] - i n \Gamma +m_0 ,
\label{eq:mom-space-action}
\end{eqnarray}
where a new definition of $\Gamma$ must be now used, as explained in the
next Section. After Fourier transforming to position space and then adding the
gauge interactions, this action corresponds to the expression given in
Eq.~(\ref{eq:generalized}). 

One can check that the continuum limit is now the correct one, and indeed
the leading term for small $p$ of (\ref{eq:mom-space-action}) is
$i\slash{p}$, as it should be. What has happened is that $\gamma'_\mu$ terms 
like the ones in Eq.~(\ref{eq:wrong-cl}), which spoiled the continuum limit
for the action (\ref{eq:action2}), are now exactly compensated by the new
$\cot \alpha_\mu$ terms that have been introduced in the modified naive
actions.

\section{Dirac matrices}
\label{sec:dirac}

The generalized definition of $\Gamma$ that, when combined with the sum of the
modified naive actions, builds a minimally doubled action is
\begin{equation}
\Gamma = \frac{1}{n} \sum_\mu \,
  \frac{1-\cos \alpha_\mu}{\sin \alpha_\mu} \, \gamma_\mu
       = \frac{1}{n} \sum_\mu \,
  \frac{1-\cos \alpha_\mu}{\sin \alpha_\mu} \, \gamma'_\mu \  ;
\quad
n = \sqrt{ \, \sum_\mu \frac{(1-\cos \alpha_\mu)^2}{\sin^2 \alpha_\mu} } \  .
\label{eq:biggamma}
\end{equation} 
With this choice the action (\ref{eq:mom-space-action}) has always two zeros,
located at the origin and at $\alpha_\mu$. If the components of $\alpha_\mu$
become large, other zeros can however appear, as discussed in the next Section.

The matrix $\Gamma$ encodes the generic direction of hypercubic breaking that
is now possible to choose. Note that one can also write it as
\begin{equation}
\Gamma = \frac{1}{n} \sum_\mu \, \gamma_\mu \, \tan (\alpha_\mu/2) ,
\end{equation} 
from which it is perhaps easier to see that there is a one-to-one correspondence
between $\Gamma$ and the direction of hypercubic breaking.

The normalization of the hermitian matrix $\Gamma$ in (\ref{eq:biggamma})
is such that $\Gamma^2=1$, and so this matrix is also unitary. We can then
observe that also the two modified naive actions out of which the action
(\ref{eq:mom-space-action}) was built are physically equivalent. Indeed if,
along the lines of \cite{Creutz:2008sr}, we consider the unitary transformation
\begin{eqnarray}
           \psi (x) &\to & e^{-i\alpha_\mu x_\mu}\,\,\Gamma\,\,\psi (x) \\
\overline{\psi} (x) &\to & e^{i\alpha_\mu x_\mu}\,\,\overline{\psi} (x)\,\,\Gamma ,
\end{eqnarray}
in momentum space the corresponding effect is given by the substitutions
$\sin (p_\mu) \to \sin (p_\mu+\alpha_\mu)$ and 
$\cos (p_\mu) \to \cos (p_\mu+\alpha_\mu)$. Thus, under this unitary
transformation the first modified naive action goes exactly into the second
one.  

An important consequence of this equivalence is that the relation
$\gamma'_\mu = \Gamma \gamma_\mu \Gamma$ of the standard Bori\c{c}i-Creutz action
is still valid, even though now the explicit expressions of the $\gamma'_\mu$
matrices depend on the choice of $\alpha_\mu$.\,\footnote{This equivalence can
be seen from the fact that the unitary transformation brings the first zero
onto the second one, and so
$\overline{\psi} \gamma_\mu \psi \to 
\overline{\psi} \Gamma \gamma_\mu \Gamma \psi
= \overline{\psi} \gamma'_\mu \psi$.}
Moreover, from this relation (and together with $\Gamma^2=1$) the equivalence
of the two definitions of $\Gamma$ in Eq.~(\ref{eq:biggamma}) can be verified,
as well as that
\begin{equation}
\{ \gamma'_\mu, \gamma'_\nu \} = 
\{ \Gamma \gamma_\mu  \Gamma , \Gamma \gamma_\nu \Gamma \} =
\Gamma \{ \gamma_\mu, \gamma_\nu \} \Gamma = 2 \delta_{\mu\nu} ,
\end{equation}
which indicates that the matrices $\gamma'_\mu$ constitute a fully legitimate
set of Dirac matrices. In general they are a linear combination of the
$\gamma_\mu$, which can be expressed as
$\gamma'_\mu = \sum_\nu a_{\mu\nu} \gamma_\nu$, where $a$ is an orthogonal matrix
\cite{Borici:2007kz}. The specific values of the entries of $\gamma'_\mu$
depend on the actual location of the second zero.

It is also easy to see, using $\gamma'_\mu = \Gamma \gamma_\mu \Gamma$, that
$\{\gamma'_\mu,\gamma_5\}=0$, and from this the chiral symmetry and the
$\gamma_5$-hermiticity of the action immediately follow. Another useful
relation for the $\gamma'_\mu$ matrices is\,\footnote{This relation was
$\gamma'_\mu = \Gamma - \gamma_\mu$ for the standard Bori\c{c}i-Creutz action. 
The relation $\gamma'_\mu = \Gamma \gamma_\mu \Gamma$ remains instead unmodified
also in the generalized Bori\c{c}i-Creutz action, and it so looks as though
it could be the more fundamental of the two main ways of expressing
$\gamma'_\mu$ in terms of $\gamma_\mu$.}
\begin{equation}
\gamma'_\mu = \{\Gamma,\gamma_\mu\}\Gamma -\gamma_\mu =
\frac{2}{n} \,
\frac{1-\cos \alpha_\mu}{\sin \alpha_\mu} \, \Gamma - \gamma_\mu .
\label{eq:usefulrelation}
\end{equation}

Note that it must be $\alpha_\mu \neq 0$ and $\alpha_\mu \neq \pi$, otherwise
the two modified naive actions of Sect. \ref{sec:construction} collapse onto
each other or their sum is identically zero, and thus the construction of the
action (\ref{eq:mom-space-action}) obviously degenerates. The direction of 
hypercubic breaking can then never exactly correspond to one of the $p_\mu$
axes. Note also that writing
$\Gamma = (1/n)\,\sum_\mu \gamma_\mu \tan (\alpha_\mu/2)$ could erroneously
suggest that $\alpha_\mu=0$ might also be included, however this cannot be
a legitimate choice.

If the second zero is constrained to remain on the positive major diagonal,
i.e. $\alpha_\mu = (\alpha,\alpha,\alpha,\alpha)$, then the formula for
$\Gamma$ reduces to the simple expression of Eq.~(\ref{eq:bcbiggamma}),
the one of the standard Bori\c{c}i-Creutz action. This also happens with 
other relations like the one in (\ref{eq:usefulrelation}).

For $\alpha_\mu = (\pi/2,\pi/2,\pi/2,\pi/2)$ all above formulae reduce to those
of the standard Bori\c{c}i-Creutz action. In particular, the normalization
factor can be seen to behave like $n \to 2$ when $\alpha_\mu \to \pi/2$.

What we have generalized here is the standard Bori\c{c}i-Creutz action whose
second zero is conventionally taken at $(\pi/2,\pi/2,\pi/2,\pi/2)$,
and hence its direction of hypercubic breaking is the positive major diagonal.
However, from the second (modified) naive action one could choose any of its
other 15 zeros out of $(\pm\pi/2,\pm\pi/2,\pm\pi/2,\pm\pi/2)$ to survive at
the end in the final combination (\ref{eq:mom-space-action}). If for instance
one picks $(\pi/2,-\pi/2,\pi/2,\pi/2)$ as the second zero, then the new
direction of hypercubic breaking is a different major hypercubic diagonal,
and reflected in the new choice
$\Gamma=\frac{1}{2}\,(\gamma_1-\gamma_2+\gamma_3+\gamma_4)$.
Each of these 16 possible choices corresponds to a restriction to a
four-dimensional orthant, that is to only one sixteenth of the whole first
Brillouin zone. The generalized action (\ref{eq:mom-space-action}) that we have
derived is instead valid for all the sixteen orthants combined (except for
the $p_\mu$ axes). One can see that also the expression for $\Gamma$ given in
(\ref{eq:biggamma}) already covers this general case, and for example if
$-\pi < \alpha_2 < 0$ then the coefficient of $\gamma_2$ in (\ref{eq:biggamma})
becomes automatically negative.

It is thus in general possible to choose $\alpha_\mu$ in any location as second
zero of the action, as long as every component of it differs from $0$ or $\pi$.

However, not all possible choices of $\alpha_\mu$ in the first Brillouin zone
preserve minimal doubling. As we discuss in the next Section, in general
additional zeros can appear if some components of $\alpha_\mu$ become too
large. It can however be proven that for a large region of choices of
$\alpha_\mu$ there are indeed only two flavors.

\section{Minimal doubling}
\label{sec:minimal}

It is interesting to note that a special situation for the modified naive
actions arises when any component of $\alpha_\mu$ becomes exactly
$\alpha_\mu=2\pi/3$. Since in this case (and only in this case, up to sign
flips of the components) $\alpha_\mu$ satisfies $2\alpha_\mu-2\pi=-\alpha_\mu$,
what happens is that the $p_\mu$ component of one zero of the first modified
naive action coincides with the $p_\mu$ component of one zero of the second
modified naive action. This is the only case for which this can happen
(together with its mirror case $\alpha_\mu=-2\pi/3$).

It actually turns out that for $\alpha_\mu = (2\pi/3,2\pi/3,2\pi/3,2\pi/3)$
the generalized Bori\c{c}i-Creutz action has indeed additional zeros,
such as\,\footnote{I thank Mike Creutz for alerting me to the existence of
this additional zero.}
\begin{equation}
p_\mu = (\pi/3,\pi/3,\pi/3,-2\pi/3) .
\label{eq:thatzero}
\end{equation}
This is not entirely trivial, as for these particular values of $\alpha_\mu$
and $p_\mu$ the action is at first sight given by
\begin{equation}
D(p) = i\frac{2}{\sqrt{3}} \, \sum_{k=1}^3 (\gamma_k + \gamma'_k) - in \Gamma ,
\end{equation}
with $n=2\sqrt{3}$. However, if one now uses
$\gamma'_k = (2\sqrt{3}/n) \, \Gamma - \gamma_k$ from
Eq.~(\ref{eq:usefulrelation}), then all terms cancel out (even without
expanding $\Gamma$), and so $D(p)$ indeed vanishes. As noted before, all
nontrivial permutations of the components of this $p_\mu$ also correspond to
additional zeros of the action.\,\footnote{It might be interesting to observe
that also the action with the wrong continuum limit, Eq.~(\ref{eq:action2}),
becomes additional zeros for the same value of $\alpha = 2\pi/3$, and at the
same locations in momentum space, $p_\mu = (\pi/3,\pi/3,\pi/3,-2\pi/3)$ plus
its nontrivial permutations. The trace equations for this action, analogous
to those discussed in Appendix \ref{sec:derivations}, are
$\Tr\,(\gamma_\mu/\sin \alpha_\mu -\gamma_\nu/\sin \alpha_\nu)\,D(p) = 0$ and
$\Tr\,\Gamma\,D(p)=0$, and give respectively
\begin{equation}
\frac{\sin p_\mu -\sin (\alpha_\mu-p_\mu)}{\sin \alpha_\mu} =
\frac{\sin p_\nu -\sin (\alpha_\nu-p_\nu)}{\sin \alpha_\nu} 
\end{equation}
and
\begin{equation}
\sum_\mu \, \sin \alpha_\mu \, 
(\sin p_\mu -\sin \alpha_\mu +\sin (\alpha_\mu - p_\mu)) = 0 .
\end{equation}
It can be easily verified that the values of $\alpha_\mu$ and $p_\mu$ given
above satisfy these trace equations (in addition to the ``standard'' zeros).
This happens in spite of the fact that for the naive (not modified) actions
of Eq.~(\ref{eq:action2}) this value of $\alpha_\mu$ (and indeed any other one)
does not give rise to a special situation. It could be that this particular
additional zero at this specific location is generated by some general
underlying mechanism, to a certain extent independent of the details of this
kind of minimally doubled actions. Whether this is indeed a general feature
of all these actions remains still to be understood.}

Any zero of the action (\ref{eq:mom-space-action}) has to satisfy the trace
equations 
\begin{equation}
\sum_\mu \frac{\cos \Big(p_\mu -\alpha_\mu/2\Big)}{
\cos \big(\alpha_\mu/2\big)} = 4 , \quad
\frac{\sin \Big(p_\mu -\alpha_\mu/2\Big)}{\sin \big(\alpha_\mu/2\big)} =
\frac{\sin \Big(p_\nu -\alpha_\nu/2\Big)}{\sin \big(\alpha_\nu/2\big)} ,
\end{equation}
which are introduced and discussed in more detail in Appendix
\ref{sec:derivations}, and come out from imposing
$\Tr\,(\gamma_\mu \sin \alpha_\mu/(1 - \cos \alpha_\mu) -\gamma_\nu
\sin \alpha_\nu/(1 - \cos \alpha_\mu))\,D(p) = 0$ and
$\Tr\,\Gamma\,D(p)=0$ respectively. It is easy to verify that the values of
$\alpha_\mu$ and $p_\mu$ given above for the additional zero satisfy these
trace equations.

Two important general properties of the zeros can be straightforwardly
inferred from the trace equations.

One is that there is a complete symmetry of the zeros under permutations of
the coordinates. Hence, in the following all zeros will be meant and given
up to nontrivial permutations of $\alpha_\mu$ and $p_\mu$.

The other one follows from the fact that these equations are symmetric under
reflections of any of the coordinates axes. Then, if a certain $p_\mu$ is a zero
for the action given by $\alpha_\mu$, changing sign to one component of $p_\mu$
will automatically give a zero of the action corresponding to an $\alpha_\mu$
which has also undergone the same change of sign. An important consequence of
this is that each orthant can then be studied separately, since the distribution
patterns of the zeros is the same in every orthant, and only changes of signs
have to be taken into account. We can then restrict our considerations to
$\alpha_\mu$'s which have only positive components, that is to the first orthant
(which corresponds to generalizing, of the 15 standard Bori\c{c}i-Creutz
actions, the one where the second zero is on the positive major diagonal).
In the following discussions $0 < \alpha_\mu < \pi$ will be always understood.

With the help of these trace equations one can always check, by direct
inspection, whether or not a given $p_\mu$ is a zero of the action for a
given choice of $\alpha_\mu$. It is easy to see that these equations are
satisfied by the two ``standard'' zeros $p_\mu = (0,0,0,0)$ and
$p_\mu = (\alpha_\mu,\alpha_\mu,\alpha_\mu,\alpha_\mu)$,
and that mixed choices such as $p_\mu=(\alpha_1,0,\alpha_3,\alpha_4)$ instead
are not zeros of the action.

Thus, the generalized action has always two zeros at the locations
$p_\mu = (0,0,0,0)$ and $p_\mu = (\alpha_\mu,\alpha_\mu,\alpha_\mu,\alpha_\mu)$.
For certain regions in the space of $\alpha_\mu$ however additional zeros can
arise, leading so to the loss of minimal doubling.

We start to look at the situation where $\alpha_\mu$ lies on the positive major
diagonal, i.e. $\alpha_\mu=(\alpha,\alpha,\alpha,\alpha)$. Then the trace
equations for the zeros become much simpler, and one can solve them
analytically along the entire length of the diagonal, and in this way see where
there are extra zeros and give an explicit expression for their locations.
The detailed derivations of the following results can be found in Appendix
\ref{sec:derivations}.

One thing that can be easily proven is that with this restriction on
$\alpha_\mu$ no extra zero of the form $p_\mu=(p,p,p,p)$ can exist, because
any zero of this form satisfies the simple equation
\begin{equation}
D(p) = \cos(p-\alpha) + \cos(p) - \cos(\alpha) - 1 = 0 ,
\end{equation}
which has no additional solutions besides the ``standard'' ones given by
$p=0$ and $p=\alpha$. So, wherever there are additional doublers on this
diagonal of $\alpha_\mu$, these cannot have all components of $p_\mu$ equal.

For $\alpha_\mu=(\alpha,\alpha,\alpha,\alpha)$ one can prove (see Appendix
\ref{sec:derivations}) that if one chooses $\alpha < 2\pi/3$ there cannot be
additional zeros, and thus minimal doubling is preserved. On the other hand,
when $\alpha \ge 2\pi/3$ additional doublers do appear, and they can be
expressed (as solution of Eqs.~(\ref{eq:thatcase})) as
\begin{equation}
p_\mu = \Big(\alpha/2 + \eta_+,\alpha/2 + \eta_+,
            \alpha/2 + \eta_+,\alpha/2 + \eta_-\Big), \quad
       \eta_\pm = \arccos~(\pm 2 \cos \alpha/2) .
\label{eq:bifurcation}
\end{equation}
This gives 2 solutions for each choice of $\alpha$ (up to nontrivial
permutations of the components), which become more and more distant from each
other as the value of $\cos \alpha/2$ decreases towards $0$, where they
coalesce on the standard zeros. At the other end of the range, i.e. for
$\alpha = 2\pi/3$, the two solutions become degenerate, and correspond to the
additional zero (\ref{eq:thatzero}).

In the general case where $\alpha_\mu$ is not on a major hypercubic diagonal
it is difficult to obtain exact solutions to the trace equations. However
one can prove that minimal doubling can be guaranteed if the components of
$\alpha_\mu$ do not become too large. A uniform bound for all components
is provided by
\begin{equation}
\cos \big(\alpha_\mu/2\big) \ge \frac{3}{5},
\label{eq:ubound}
\end{equation}
which corresponds to $\alpha_\mu \le 0.590334\,\pi \sim 106.26^o$.
When this condition is true no other zeros can appear in the action.
Its derivation is given in Appendix \ref{sec:derivations}.

One can also see, by direct inspection of the trace equations, that there are
extra zeros for the actions defined by\,\footnote{Note that it must be
$\delta > 0$, and that we cannot take here $\delta = 0$, because this
corresponds to $\alpha_4=0$, in which case, as remarked in the previous
Section, the construction of the action (\ref{eq:mom-space-action})
degenerates. Note also that for $\delta=1/2$ one obtains again the additional
zero (\ref{eq:thatzero}).}
\begin{equation}
\cos \big(\alpha_\mu/2\big) = \left( \frac{3-3\delta}{5-4\delta} \, ,
\frac{3-3\delta}{5-4\delta} \ , \frac{3-3\delta}{5-4\delta} \ , 1 - \delta
\right) ,
\end{equation}
and these zeros are given by
\begin{equation}
\cos \Big(p_\mu -\alpha_\mu/2\Big) = (1,1,1,-1) .
\end{equation}
If one takes $\delta$ to be very small, the existence of these zeros shows
that it is not possible to further improve the uniform bound (\ref{eq:ubound})
given above. 

It can also be demonstrated that when for all components
\begin{equation}
\cos \big(\alpha_\mu/2\big) \le \frac{1}{2},
\end{equation}
then minimal doubling is surely lost, that is extra zeros always appear and 
one is in another branch. The derivation of this statement can also be found
in Appendix \ref{sec:derivations}, and this region corresponds to taking
$\alpha_\mu \ge 2\,\pi/3$.

Somewhere inbetween the two uniform bounds established above,
$\cos \big(\alpha_\mu/2\big) \ge 3/5$ and $\cos \big(\alpha_\mu/2\big) \le 1/2$,
there should be a 3-dimensional surface of demarcation between the domain of
minimal doubling and the branches which contain more doublers. Note that the
extra zeros just given above lie infinitesimally close to the bound for the
minimal doubling region, and that conversely (as found at the beginning of this
Section) when one moves on the major diagonal towards the origin starting
from the additional zero for the action
$\alpha_\mu = (2\pi/3,2\pi/3,2\pi/3,2\pi/3)$, which lies exactly at the bound
$\cos \big(\alpha_\mu/2\big) = 1/2$, minimal doubling is immediately restored.
It also seems from these results that the region of minimally doubling
contracts in the directions where the differences between the components of
$\alpha_\mu$ are large, whereas when all components of $\alpha_\mu$ are equal
then minimal doubling can be preserved also for larger values of them. It is
possible that the boundary of the minimal doubling domain will have a
nontrivial and complicated shape.

It is not easy to derive a general solution of the trace equations when
$\alpha_\mu$ is located outside a major hypercubic diagonal. However it is
possible to decouple them and so to write an equation for a single component
of $p_\mu$, which could turn out to be useful in other contexts. Indeed, by
combining Eq.~(\ref{eq:trbiggamma}) with Eq.~(\ref{eq:trmumnu}) one can obtain
an equation for one component $\cos \Big(p_\sigma -\alpha_\sigma/2\Big)$,
\begin{equation}
\frac{\cos \Big(p_\sigma -\alpha_\sigma/2\Big)}{\cos \big(\alpha_\sigma/2\big)} +
\sum_{\rho\neq\sigma} (\pm) \frac{1}{\cos \big(\alpha_\rho/2\big)}
\sqrt{1-\Big(1-\cos^2 \big(\alpha_\rho/2\big)\Big) \, \frac{1 - \cos^2
\Big(p_\sigma -\alpha_\sigma/2\Big)}{1-\cos^2 \big(\alpha_\sigma/2\big)}} = 4 ,
\label{eq:ntrbiggamma}
\end{equation}
where $\pm$ means that for each $\rho$ component one can take either the
positive or the negative square root. Any component $\mu$ of a momentum $p_\mu$
which is a zero of the action must then satisfy one of these 9 possible
equations.

One can further explore the whole space spanned by $\alpha_\mu$, even though
at this stage we have already a general picture of the situation which is
sufficient for first nonperturbative investigations. After all, the values of
$\alpha_\mu$ for which $\cos \big(\alpha_\mu/2\big) = 3/5$ and
$\cos \big(\alpha_\mu/2\big) = 1/2$ do not lie too distant from each other.
Of course, if needed one can always immediately check from the trace equations,
by direct inspection, whether a certain momentum $p_\mu$ is a zero of the
action or not.

Moreover, one should keep in mind that the numbers that we have given in this
Section are the result of tree-level considerations, and hence the actual
surfaces of demarcation between the regions of minimal doubling and those that
contain additional doublers may be slightly different after all interactions
have been taken into account.

\section{Outlook}
\label{sec:outlook}

In this article we have presented new minimally doubled actions of the 
Bori\c{c}i-Creutz type, with which one is able to put the second zero at any
location in momentum space (except on the $p_\mu$ axes). Minimal doubling
is preserved provided the components of $\alpha_\mu$ do not become too large.
We have indeed proven that there are always only two zeros in the first
orthant when for all components $\cos \big(\alpha_\mu/2\big) \ge 3/5$, while if
for all components $\cos \big(\alpha_\mu/2\big) \le 1/2$ then surely there are
at least four zeros. Exact solutions can be given over the whole length of the
positive major diagonal, and in this case minimal doubling is preserved also
for larger values of the components of $\alpha_\mu$, that is when
$\cos \big(\alpha_\mu/2\big) > 1/2$. This pattern is exactly repeated in
the 15 other orthants.

Recently new minimally doubled actions of a still different kind
\cite{Capitani:2013zta} were also proposed, which allow to put the second zero
anywhere on one of the $p_\mu$ axes. Thus, with all these new actions at hand,
one can now choose to put the second zero at any position in a region which 
covers a large part of the first Brillouin zone.

It seems not possible to go continuously from one kind of actions to the other
one. For example, if one takes
$\alpha_\mu=(\epsilon,\epsilon,\epsilon,\pi-\epsilon)$ in the generalized
Bori\c{c}i-Creutz action (\ref{eq:mom-space-action}), and then lets
$\epsilon \to 0$, a singularity of the action is encountered, and one cannot
connect smoothly to the standard Karsten-Wilczek action. Even taking just
$\alpha_\mu=(\epsilon,\epsilon,\epsilon,-2\alpha)$ and letting $\epsilon \to 0$
while $\alpha$ is kept constant, produces a singularity of the action before
the position of the second zero of the generalized Karsten-Wilczek actions
can be reached. This should perhaps be not too surprising, as the mechanism
out of which one can remain with only two zeros is substantially different
for these two kinds of actions. We also note that for these generalized
Bori\c{c}i-Creutz fermions the (4-dimensional) distance can be increased
up to $4\pi/3$ (on the major hypercubic diagonals) while still preserving
minimal doubling, and this is larger than the maximum distance possible with
(even generalized) Karsten-Wilczek fermions.

There is also another difference, if we look at the bare actions.
For Karsten-Wilczek fermions, the generalized case \cite{Capitani:2013zta}
included, $P$ is a conserved symmetry, but $T$ and $C$ are violated
\cite{Bedaque:2008xs}, whereas for standard Bori\c{c}i-Creutz fermions
as well as their generalization presented here, also $P$ is violated.
A violation of $C$, which occurs for all these bare actions, has among others
the consequence that the masses of the $\pi^+$ and $\pi^-$ are not equal.
This seems hard to escape, because $PT$ can never be a symmetry of any
minimally doubled action, as was explained in \cite{Bedaque:2008xs}.
Of course $C$ can be restored at the end and consequently the masses of the
$\pi^+$ and $\pi^-$ become again equal, once one has properly tuned the required
counterterms and thus constructed the properly renormalized action to be
eventually used in Monte Carlo simulations.  

For Karsten-Wilczek fermions the counterterms contain factors like
$\delta_{\mu 4}$. For standard Bori\c{c}i-Creutz fermions the characteristic
feature of the counterterms is the appearance of sums involving only one
Lorentz index, $\sum_\mu f_\mu$, which mirrors $2\Gamma=\sum_\mu \gamma_\mu$
\cite{Capitani:2010nn}. In the case of the generalized Bori\c{c}i-Creutz
fermions presented here, the sums over only one Lorentz index must be of the
form $\sum_\mu f_\mu (1-\cos \alpha_\mu)/\sin \alpha_\mu$, which mirrors the
generalized $\Gamma$. We expect then the fermionic counterterms to look
formally like the ones required for standard Bori\c{c}i-Creutz fermions,
\begin{equation}
\overline{\psi}\,\Gamma \sum_\mu D_\mu \psi, \qquad
\frac{1}{a} \,\overline{\psi}(x)\,\Gamma\,\psi(x) ,
\end{equation}
where of course the explicit expressions now depend on the actual choice of
$\alpha_\mu$. The gluonic counterterm will also contain information about the
special direction, and we expect it to be a fixed linear combination of the
components of the plaquette which reflects the specific direction of hypercubic
breaking, with only one overall coefficient to be tuned:
\begin{equation}
\sum_{\mu\nu\rho} 
\frac{1-\cos \alpha_\mu}{\sin \alpha_\mu} \,
\frac{1-\cos \alpha_\nu}{\sin \alpha_\nu} \,
\Tr\, F_{\mu\rho}(x) \, F_{\rho\nu}(x) . 
\end{equation}
Note that a further splitting of the index $\rho$ into two independent sums
gives identically zero, because of the antisymmetry of the field-strength
tensor.

It is possible that special choices of $\alpha_\mu$ can result in a reduction
of the number of counterterms, as it has occurred in the case of generalized
Karsten-Wilczek fermions \cite{Capitani:2013zta}. One should investigate
the renormalization properties of this action and consequently pick up
the best choice of $\alpha_\mu$ for Monte Carlo simulations. The hope is that
for some special value of $\alpha_\mu$ the corresponding action would require
no counterterms at all.

It appears difficult to insert more parameters in these actions of the
Bori\c{c}i-Creutz type, beyond $\alpha_\mu$. A parameter like the $\lambda$
of the generalized Karsten-Wilczek case cannot be introduced, as this is very
peculiar to that kind of minimally doubled fermions, in which the mechanism
of minimal doubling is different from the one of the Bori\c{c}i-Creutz action.
But even with $\alpha_\mu$ alone, we can vary at will not only the distance
between the two zeros, but also the direction of hypercubic breaking.
The four components of $\alpha_\mu$ can be chosen independently, and it
could be that one can regard $\alpha_\mu$ as four independent parameters
(in this sense, many more than for the generalized Karsten-Wilczek fermions
studied in \cite{Capitani:2013zta}). If the four components of $\alpha_\mu$
really behave completely independently under renormalization, this could
increase the possibility for finding some particular choice of parameters
which realizes a minimally doubled action with no counterterms.

\section*{Acknowledgments}

I have enjoyed and profited from many interesting conversations with
Mike Creutz, whom I warmly thank.

\appendix

\section{Derivations of minimal doubling}
\label{sec:derivations}

In this Appendix we provide some derivations regarding the properties 
of the zeros of the action and the possible appearance in specific domains of
extra doublers which destroy minimal doubling.

\subsection{Trace equations}

We need the basic traces
\begin{equation}
\Tr\, \Gamma \gamma_\mu = \Tr\, \Gamma \gamma'_\mu = 
\frac{4}{n} \, \frac{1-\cos \alpha_\mu}{\sin \alpha_\mu} 
\end{equation}
and
\begin{equation}
\Tr\, \gamma_\mu \gamma'_\nu = - 4 \, \delta_{\mu\nu} + \frac{8}{n^2} \,
\frac{1-\cos \alpha_\mu}{\sin \alpha_\mu} \,
\frac{1-\cos \alpha_\nu}{\sin \alpha_\nu} .
\end{equation}
We start by computing
\begin{eqnarray}
\Tr\, \gamma_\mu \, D(p) &=& 4i \,\bigg\{
\sin p_\mu + \cos p_\mu \,\frac{\cos \alpha_\mu -1}{\sin \alpha_\mu} 
\label{eq:firsttrace} \\
&& \qquad + \frac{1-\cos \alpha_\mu}{\sin \alpha_\mu} \,
\bigg( \frac{2}{n^2} \sum_\rho \frac{1-\cos \alpha_\rho}{\sin^2 \alpha_\rho}
\, \Big( \cos p_\rho - \cos \alpha_\rho \Big) -1 \bigg) \bigg\} \ , \nonumber
\end{eqnarray}
and also the computation of $\Tr\, \gamma'_\mu D(p)$ gives a similar result of
this kind. Then one can see that it is more useful to consider
\begin{equation}
   \Tr\, \Bigg( \frac{\sin \alpha_\mu}{1-\cos \alpha_\mu} \, \gamma_\mu 
      - \frac{\sin \alpha_\nu}{1-\cos \alpha_\nu} \, \gamma_\nu \Bigg) \, D(p)
= - \Tr\, \Bigg( \frac{\sin \alpha_\mu}{1-\cos \alpha_\mu} \, \gamma'_\mu 
      - \frac{\sin \alpha_\nu}{1-\cos \alpha_\nu} \, \gamma'_\nu \Bigg) \, D(p) .
\end{equation}
Then, from (\ref{eq:firsttrace}) we can easily obtain the simple result
\begin{equation}
\Tr\, \Bigg( \frac{\sin \alpha_\mu}{1-\cos \alpha_\mu} \, \gamma_\mu 
  - \frac{\sin \alpha_\nu}{1-\cos \alpha_\nu} \, \gamma_\nu \Bigg) \, D(p)
= 4i \, \Bigg(
  \frac{\sin \Big(p_\mu -\alpha_\mu/2\Big)}{\sin \big(\alpha_\mu/2\big)} -
  \frac{\sin \Big(p_\nu -\alpha_\nu/2\Big)}{\sin \big(\alpha_\nu/2\big)} \Bigg) .
\end{equation}
If $p$ is a zero of the action, then these traces must also vanish, and
the above result implies
\begin{equation}
\frac{\sin \Big(p_\mu -\alpha_\mu/2\Big)}{\sin \big(\alpha_\mu/2\big)} =
\frac{\sin \Big(p_\nu -\alpha_\nu/2\Big)}{\sin \big(\alpha_\nu/2\big)} .
\label{eq:trmumnu}
\end{equation}
These equalities (valid for all directions) are obviously satisfied by the 
two zeros of the action, $p_\mu = 0$ and $p_\mu = \alpha_\mu$. They also
prohibit mixed choices like $p_\mu=(\alpha_1,0,\alpha_3,\alpha_4)$ to be a zero.

We next compute 
\begin{equation}
\Tr\, \Gamma \, D(p) = \frac{4i}{n} \, \sum_\mu \, 
\frac{1-\cos \alpha_\mu}{\sin \alpha_\mu} \, \bigg( \sin p_\mu + 
\cos p_\mu \,\frac{\cos \alpha_\mu +1}{\sin \alpha_\mu} 
-2 \,\frac{\cos \alpha_\mu}{\sin \alpha_\mu} \bigg) -4in .
\end{equation}
Equating to zero this result we arrive, after some rearrangements, to
\begin{equation}
\sum_\mu \frac{\cos \Big(p_\mu -\alpha_\mu/2\Big)}{
\cos \big(\alpha_\mu/2\big)} = 4 .
\label{eq:trbiggamma}
\end{equation}
This is again satisfied by the two standard zeros of the action, and for them
all four terms in the sum are equal to one.

As we already noted in Sect. \ref{sec:minimal}, from the trace equations 
one can clearly see that the zeros are symmetric under permutations of the
coordinates and under reflections of any of the coordinates axes.
In particular, this allows us to restrict all the following discussions to 
the first orthant, that is $0 < \alpha_\mu/2 < \pi/2$, which also implies that
both $\sin \big(\alpha_\mu/2\big)$ and $\cos \big(\alpha_\mu/2\big)$
are always positive.

\subsection{On a major diagonal}

In the particular case in which $\alpha_\mu$ lies on a major hypercubic
diagonal, the trace equations become much simpler and they can be solved for
the entire length of the diagonal. Indeed, when
$\alpha_\mu=(\alpha,\alpha,\alpha,\alpha)$ Eq.~(\ref{eq:trmumnu}) becomes
\begin{equation}
\sin \Big(p_\mu -\alpha/2\Big) = \sin \Big(p_\nu -\alpha/2\Big),
\label{eq:eqsineq}
\end{equation}
which implies
\begin{equation}
\cos \Big(p_\mu -\alpha/2\Big) = \pm \cos \Big(p_\nu -\alpha/2\Big),
\label{eq:eqcospm}
\end{equation}
and Eq.~(\ref{eq:trbiggamma}) becomes
\begin{equation}
\sum_\mu \cos \Big(p_\mu -\alpha/2\Big) = 4 \cos \big(\alpha/2\big) .
\label{eq:eqcos4}
\end{equation}

In the first orthant it is always $\alpha>0$. Now, all cosines on the left hand
side of the last equation must have, because of Eq.~(\ref{eq:eqcospm}), the
same absolute value. Then either they are all of the same sign (in this case,
positive), or one of them has opposite sign (in this case, negative) with
respect to the other three. The solutions of the first case give the standard
zeros, and those of the second case (if they exist) the additional doublers.

It is easy to prove that the second possibility can be excluded if 
$\alpha < 2\pi/3$, and thus for this domain of $\alpha$ one maintains
minimal doubling. Indeed, Eq.~(\ref{eq:eqcos4}) with the constraint
$\cos \alpha/2 > 1/2$ means that
\begin{equation}
\sum_\mu \cos \Big(p_\mu -\alpha/2\Big) > 2 ,
\label{eq:diseqcos4}
\end{equation}
and this implies that, since all $\cos \Big(p_\mu -\alpha/2\Big)$ must have
the same absolute value, then none of them can be negative, because otherwise
the result of the sum would be smaller than 2. Thus, all cosines must be equal
and have the same sign, and Eq.~(\ref{eq:eqcos4}) becomes
\begin{equation}
\cos \Big(p_\mu -\alpha/2\Big) = \cos \alpha/2 ,
\end{equation}
which can only be satisfied by the two ``standard'' zeros $p_\mu = (0,0,0,0)$
and $p_\mu = (\alpha,\alpha,\alpha,\alpha)$. It is also easy to see that mixed
choices such as $p_\mu=(\alpha,0,\alpha,\alpha)$ cannot be zeros of the action,
because they do not satisfy Eq.~(\ref{eq:eqsineq}).

If, on the other hand, we are in the region $\cos \alpha/2 \le 1/2$, then
Eq.~(\ref{eq:eqcos4}) can still have solutions for $p_\mu$ even when one of the
cosines in its left hand side is negative, such as
\begin{equation}
\cos \Big(p_\mu -\alpha/2\Big) = \left( 2 \cos \alpha/2,
2 \cos \alpha/2, 2 \cos \alpha/2, - 2 \cos \alpha/2 \right) .
\label{eq:thatcase}
\end{equation}
These are then extra zeros. It is easy to see that for $\alpha=2\pi/3$
there is only one solution, the one that we have already discussed in Sect.
\ref{sec:minimal}, $p_\mu = (\pi/3,\pi/3,\pi/3,-2\pi/3)$ plus its 3 nontrivial
permutations (see Eq.~(\ref{eq:thatzero})). When $\alpha > 2\pi/3$ this zero
bifurcates, and the distance between the two solutions grows with $\alpha$.
These are the solutions whose explicit expressions are given in
Eq.~(\ref{eq:bifurcation}) of Sect. \ref{sec:minimal}. 

We want now to understand under what conditions other zeros besides the
``standard'' ones can appear when $\alpha_\mu$ is not restricted to be
on a major hypercubic diagonal, and establish some bounds for the regions of
minimal doubling.

\subsection{Uniform bounds in the general case}

The trace equations for the sines, (\ref{eq:trmumnu}), constrain the ratios of
sines to have the same value
\begin{equation}
R = \frac{\sin \Big(p_\mu -\alpha_\mu/2\Big)}{\sin \big(\alpha_\mu/2\big)} 
\end{equation}
in all four spacetime directions. The ratios of cosines can then be expressed
for any given choice of $R$ as
\begin{equation}
\frac{\cos \Big(p_\mu -\alpha_\mu/2\Big)}{\cos \big(\alpha_\mu/2\big)} =
\pm \sqrt{R^2 + \frac{1-R^2}{\cos^2 \big(\alpha_\mu/2\big)}} .
\label{eq:ratiocosr}
\end{equation}
It is easy to see from the trace equations for the cosines,
(\ref{eq:trbiggamma}), that it is not possible for these ratios of cosines
to be greater than 1 at the same time for all four spacetime directions,
and neither is possible that they are all smaller than 1 in the four directions.
The case in which these ratios of cosines are all equal to 1 gives the
``standard'' zeros. Thus, for the action to have extra zeros in addition to the
``standard'' ones there must be at least one direction in which the ratio of
the cosines in Eq.~(\ref{eq:trbiggamma}) is (strictly) greater than 1, and
also at least one direction in which the ratio is instead (strictly) smaller
than 1.

Now, the fact that at least one ratio of cosines must be greater than 1 implies
that $R^2<1$ (as can be seen for instance from Eq.~(\ref{eq:ratiocosr})).
But this in turn constrains those ratio of cosines that are smaller than 1
in the cosine equation (\ref{eq:trbiggamma}) to the much stronger requirement
of being actually smaller than $-1$. This observation is crucial for the
following.

Let us at this point try to establish a uniform bound of the kind
\begin{equation}
\cos \big(\alpha_\mu/2\big) \ge C 
\end{equation}
for the regions where extra zeros cannot appear (so that minimal doubling
is preserved). Since, as we noted before, for at least one direction the ratio
of cosines in the cosine equation must be (strictly) smaller than 1, and then
actually negative and (strictly) smaller than $-1$, let us consider the case
where this occurs for only one direction (while for the other three directions
is instead positive and strictly greater than 1). Then, $-1$ is the strict
upper bound for this one negative term in the left hand side of the cosine
equation (\ref{eq:trbiggamma}), while Eq.~(\ref{eq:ratiocosr}) allows us to set
an upper bound, which depends on $C$, for the sum of the other three terms.

Now, extra zeros can be certainly excluded if the left hand side of the cosine
equation in this configuration remains always (strictly) smaller than 4, and
this is the case if the inequality
\begin{equation}
3 \, \sqrt{R^2 + \frac{1-R^2}{C^2}} \le 5 
\end{equation}
holds. This is $C^2 \ge (R^2-1)/(R^2-25/9)$, which gives the uniform bound
$C \ge 3/5$ (attained when $R=0$). Thus, when for all directions
\begin{equation}
\cos \big(\alpha_\mu/2\big) \ge \frac{3}{5} ,
\end{equation}
which amounts to $\alpha_\mu \le 0.590334\,\pi \sim 106.26^o$, minimal doubling
is guaranteed. It is easy to see that the remaining cases in which the ratios of
cosines are negative for more than one direction cannot improve this bound.

We next prove that when for all directions the uniform bound
\begin{equation}
\cos \big(\alpha_\mu/2\big) \le \frac{1}{2}
\label{eq:boundextra}
\end{equation}
is satisfied, then there are always extra zeros, and minimal doubling is lost.
For this we will also use again some considerations from the previous
subsection.

Using the expression (\ref{eq:ratiocosr}) (which already contains the
information from Eq. (\ref{eq:trmumnu})), the cosine equation
(\ref{eq:trbiggamma}) can be rewritten as 
\begin{equation}
  \sqrt{R^2 + \frac{1-R^2}{\cos^2 \big(\alpha_1/2\big)}}
+ \sqrt{R^2 + \frac{1-R^2}{\cos^2 \big(\alpha_2/2\big)}}
+ \sqrt{R^2 + \frac{1-R^2}{\cos^2 \big(\alpha_3/2\big)}}
- \sqrt{R^2 + \frac{1-R^2}{\cos^2 \big(\alpha_4/2\big)}} = 4 ,
\label{eq:sqrootcos}
\end{equation}
where we consider one of the possible choices of signs in front of the square
roots which can give extra zeros (as discussed in the previous subsection).

Thanks to the symmetry under permutations we can always assume that
$\cos \big(\alpha_3/2\big) \le \cos \big(\alpha_4/2\big)$.\,\footnote{Otherwise
one has to consider a different equation, similar to (\ref{eq:sqrootcos}) but
where the negative sign is in front of another square root instead of the
fourth one.} 

We know from the previous subsection that when there are extra zeros it must
also be $R^2 < 1$. When $R^2$ increases from 0 to 1 the value of each square
root decreases monotonically, from its possible maximum
$1/\cos \big(\alpha_\mu/2\big)$ to its possible minimum 1. Moreover, because
$\cos \big(\alpha_3/2\big) \le \cos \big(\alpha_4/2\big)$ the difference
of the third and fourth square roots must be nonnegative. 

When $R=0$ the sum of the first and second square roots assumes its maximum
possible value, which because of (\ref{eq:boundextra}) has to be at least 4.
Since the difference of the third and fourth square roots is nonnegative,
the left hand side of (\ref{eq:sqrootcos}) gives then a result which is always
greater or equal to 4, when $R=0$.
  
When $R^2=1$, the result of the left hand side of (\ref{eq:sqrootcos}) assumes
its possible minimum, which is exactly 2 (whatever the choice of $\alpha_\mu$). 

Because of the monotonicity of each square root, and of the fourth square
root being always smaller in absolute value than the third one (or at most
equal to it), the left hand side of (\ref{eq:sqrootcos}) will decrease
monotonically when $R^2$ increases from $R^2=0$ (where its value it at least 4)
towards $R^2=1$ (where its value is 2). Then there is always a value of $R^2$
for which the left hand side will become exactly equal to 4. This demonstrates
that extra zeros always exist if the bound (\ref{eq:boundextra}) holds.

\end{document}